\documentclass[aps,prd,10pt,notitlepage,nofootinbib,superscriptaddress]{revtex4-1}

\usepackage[utf8]{inputenc}
\usepackage{amsmath}
\usepackage{amssymb}
\usepackage{amsfonts}
\usepackage{newtxtext,newtxmath}
\usepackage{bm}
\usepackage{graphicx}
\usepackage[usenames,dvipsnames]{xcolor}
\usepackage{color}
\usepackage[colorlinks=true,linkcolor=blue,urlcolor=blue,citecolor=blue]{hyperref}

\usepackage{slashed}
\usepackage[english]{babel}
\usepackage{dcolumn}
\usepackage{blindtext}
\usepackage{epsfig}
\usepackage{pifont}
\usepackage{dsfont,mathrsfs}
\usepackage{cancel}
\usepackage{bigints}
\usepackage{accents}
\usepackage{soul}
\usepackage{multirow}
\usepackage{simpler-wick}

\newcommand{\nn}{\nonumber}
\newcommand{\ket}[1]{\left|#1\right\rangle}

\newcommand{\expectationvalue}[3]{\langle#1|#2|#3\rangle}
\newcommand{\Expectationvalue}[3]{\left\langle#1\left|#2\right|#3\right\rangle}
\newcommand{\ensembleaverage}[1]{\left\langle#1\right\rangle}

\newcommand{\Ensembleaverage}[1]{\langle#1\rangle}

\newcommand{\MB}[1]{\left|#1\right|}
\newcommand{\FB}[1]{\left(#1\right)}

\newcommand{\SB}[1]{\left\{#1\right\}}
\newcommand{\TB}[1]{\left[#1\right]}

\newcommand{\mcT}{\mathcal{T}}

\newcommand{\scrL}{\mathscr{L}}

\newcommand{\scrD}{\mathscr{D}}

\newcommand{\munu}{{\mu\nu}}

\newcommand{\Tr}{\text{Tr}}

\newcommand{\kpll}{k_\parallel}

\newcommand{\ppll}{p_\parallel}

\newcommand{\kper}{k_\perp}
\newcommand{\pper}{p_\perp}

\newcommand{\gpll}{g_\parallel}
\newcommand{\gper}{g_\perp}

\newcommand{\del}{\partial}
\newcommand{\identity}{\mathds{1}}

\newcommand{\ku}{\kappa_\text{u}}
\newcommand{\kd}{\kappa_\text{d}}

\newcommand{\psibar}{\overline{\psi}}
\newcommand{\qbar}{\overline{q}}
\newcommand{\ubar}{\overline{u}}

\newcommand{\ptilde}{\tilde{p}}

\newcommand{\tu}{\text{u}}
\newcommand{\td}{\text{d}}

\newcommand{\ptd}{\tilde{p}}
\newcommand{\ptdvec}{\tilde{\bm{p}}}

\newcommand{\ktdvec}{\tilde{\bm{k}}}
\newcommand{\wptd}{\tilde{\omega}_p}
\newcommand{\wktd}{\tilde{\omega}_k}

\newcommand{\wl}{\omega_{lfs_k}}
\newcommand{\wn}{\omega_{nfs_p}}

\begin{document}
\title{Thermomagnetic modification of the anomalous magnetic moment of quarks using the NJL model}

\author{Snigdha Ghosh}
\email{snigdha.physics@gmail.com}
\affiliation{Government General Degree College Kharagpur-II, Madpur, Paschim Medinipur - 721149, West Bengal, India}

\author{Nilanjan Chaudhuri}
\email{sovon.nilanjan@gmail.com}
\affiliation{Variable Energy Cyclotron Centre, 1/AF Bidhannagar, Kolkata - 700064, India}
\affiliation{Homi Bhabha National Institute, Training School Complex, Anushaktinagar, Mumbai - 400085, India}
%
%
\author{ Pradip Roy}
\email{pradipk.roy@saha.ac.in}
\affiliation{Saha Institute of Nuclear Physics, 1/AF Bidhannagar, Kolkata - 700064, India}
\affiliation{Homi Bhabha National Institute, Training School Complex, Anushaktinagar, Mumbai - 400085, India}
\author{Sourav Sarkar}
\email{sourav@vecc.gov.in}
\affiliation{Variable Energy Cyclotron Centre, 1/AF Bidhannagar, Kolkata - 700064, India}
\affiliation{Homi Bhabha National Institute, Training School Complex, Anushaktinagar, Mumbai - 400085, India}

\begin{abstract}
The effective photon-quark-antiquark ($\gamma q \overline{q}$) vertex function is evaluated at finite temperature in the presence of an arbitrary external magnetic field using the two-flavor gauged Nambu--Jona-Lasinio (NJL) model in the mean field approximation. The lowest order diagram contributing to the magnetic form factor and the anomalous magnetic moment (AMM) of the quarks is calculated at finite temperature and external magnetic field using the imaginary time formalism of finite temperature field theory and the Schwinger proper time formalism. The Schwinger propagator including all the Landau levels with non-zero AMM of the dressed quarks is considered while calculating the loop diagram. Using sharp as well as smooth three momentum cutoff, we regularize the UV divergences arising from the vertex function and the parameters of our model are chosen to reproduce the well known phenomenological quantities at zero temperature and zero magnetic field, such as pion-decay constant ($f_\pi$), vacuum quark condensate, vacuum pion mass ($m_\pi$) as well as the magnetic moments of proton and neutron. We then study the temperature and magnetic field dependence of the AMM and constituent mass of the quark. We found that, the AMM as well as the constituent quark mass are large at the chiral symmetry broken phase in the low temperature region. Around the pseudo-chiral phase transition they decrease rapidly and at high temperatures both of them approach vanishingly small values in the symmetry restored phase. 
\end{abstract}

\maketitle
%
\section{INTRODUCTION}\label{sec.intro}
The influence of an external magnetic field on the vacuum structure of quantum chromodynamics (QCD) and its modifications at finite temperature and/or chemical potential can play an important role in many physical systems (see Ref.~\cite{Lect_note} for review).  For example, it is conjectured by some cosmological models that during the electroweak phase transition in the early universe, extremely strong magnetic field as high as $ \sim 10^{23}$ G might have been produced~\cite{Vachaspati,Campanelli} (note that in natural units, $ 10^{18} {\rm ~G} \approx m_\pi^2 \approx 0.02~ {\rm GeV}^2  $). The magnetic field on the surface of certain compact stars called \textit{magnetars}, is of the order of $ \sim10^{15}$ G, while in the interior it is estimated to reach  about $\sim 10^{18} $ G~\cite{Duncan,Duncan2,Lai}. Most importantly, in non-central or asymmetric heavy-ion collisions (HICs) at RHIC and LHC,  strong magnetic fields of the order of $ \sim10^{18} $ G~\cite{Kharzeev,Skokov2} or larger may be transiently generated. It is, however, predicted that the presence of a finite electrical conductivity of the hot and dense medium created during HICs can delay the decay of these time-dependent magnetic fields substantially~\cite{Tuchin,Gursoy,Tuchin2,Ajit_Sriv}. Thus, being comparable to the QCD scale \textit{i.e.} $eB\approx m_\pi^2 $, such high magnetic fields can influence substantial change in the deconfined medium of strongly interacting quarks and gluons known as the quark-gluon plasma (QGP) which is supposed to be created in such HICs. So far, a considerable amount of research has been conducted in the last few decades to understand the consequences of this background magnetic field on the QCD matter; this results in a large number of novel and interesting phenomena, such as, the Chiral Magnetic Effect (CME)~\cite{Fukushima,Kharzeev,Kharzeev2,Bali}, Magnetic Catalysis (MC)~\cite{Shovkovy,Gusynin1,Gusynin2,Gusynin3}  and Inverse Magnetic Catalysis (IMC)~\cite{Preis,Preis2} of dynamical chiral symmetry breaking which may cause significant change in the nature of electro-weak~\cite{Elmfors,Skalozub,Sadooghi_ew,Navarro}, chiral and superconducting phase transitions~\cite{Fayazbakhsh1,Fayazbakhsh2,Skokov,Fukushima2}, electromagnetically induced superconductivity and superfluidity~\cite{Chernodub1,Chernodub2} and many more.

However, a first principle analysis of the above mentioned phenomena involves a great deal of  complexities due to the large coupling  strength of QCD in the low energy regime which restricts the applicability of the perturbative analysis. One may rely on the lattice QCD (LQCD) simulations which provide one of the best strategies to overcome this problem at zero baryon density. It is also possible to extrapolate the zero baryon chemical potential results of LQCD for several thermodynamical quantities to the intermediate temperatures (comparable to the QCD scale) and low baryonic density using methods, such as, Taylor expansion~\cite{Bazavov2017} or an analytical continuation from imaginary chemical potentials~\cite{Gunther} which is relevant for highly relativistic HICs~\cite{Bazavov2,Bazavov,Bazavov2017,Gunther,Bali_lat,Bali_lat3,Sharma}. But, for examining compact stars, one has to deal with low temperature and high density extreme states, which are also expected to be explored in the upcoming CBM experiment at FAIR. The so-called sign problem in the Monte Carlo sampling restricts the accessibility of these areas of the phase diagram via the LQCD simulation. In this situation, available alternative is to work with QCD inspired effective models which possess some of the essential features of QCD and study the effects of background  magnetic field on such effective description~\cite{Andersen}. Nambu--Jona-Lasinio (NJL) model~\cite{Nambu1,Nambu2} is one such model, which is constructed respecting the global symmetries of QCD, most importantly the chiral symmetry (see Refs.~\cite{Klevansky,Hatsuda1,Vogl,Buballa} for reviews). This model has been extensively used to study phase structure of hot, dense and magnetized QCD medium~\cite{Andersen,Klevansky2,Gusynin1,Gusynin2,Gusynin3,Mao,Sadooghi1,Ruggieri,Snigdha1,Avancini,Avancini_reg,AvanciniCEP,Zhang}.

The appearance of anomalous magnetic moment (AMM) of an elementary particle, having no internal structure, due to quantum corrections in Quantum Electrodynamics (QED) is a well known phenomenon in gauge field theory.  When the Fermions are coupled to the gauge field via minimal coupling, the ordinary derivatives are modified to the covariant derivatives $\partial_\mu \rightarrow D_\mu = \partial_\mu - iQeA_\mu  $ and the Dirac equation can be recast as~\cite{Peskin:1995ev,Schwartz,Schwinger_MM}
\begin{equation}\label{key}
\FB{D^2 - g\frac{Qe}{4} F^{\mu\nu}\sigma_{\mu\nu} +m^2} \psi =0
\end{equation}
where $ \sigma_{\mu\nu}= i[\gamma_\mu, \gamma_\nu]/2 $ and $ g $ is the Land\'e g-factor.  For example, the Land\'e g-factor of the electron comes out to be $ 2 + \alpha/\pi  $ up to one-loop in QED where $ \alpha  $ is the fine structure constant. One can also consider the higher order corrections to $ g $ in power series of $ \alpha/\pi  $~\cite{Mohr} which are in excellent agreement with experimental data. In presence of a background magnetic field, the AMM modifies the mass of electron as  $ m_{\rm eff}^2 \approx m^2+ (g-2)eB/2 $  in the lowest Landau level (LLL)~\cite{Mao:2018jdo}. However, in case of massless QED, chiral symmetry breaking leads to the dynamical generation of AMM~\cite{Ferrer_prl,Ferrer_NPB}. Now, QCD being the gauge theory of strong interactions, an anomalous contribution to the magnetic moment can therefore be associated with the quarks due to strong corrections, along with the QED corrections. 
But, the non-perturbative nature of QCD forbids one to perform a first principle analytical calculation to extract the 	anomalous contribution to $ g $. Thus, one has to resort to effective models, such as NJL model with spontaneous symmetry breaking (SSB) which, in addition to the effective mass (chiral condensate), leads to the dynamical generation of the AMM in a magnetic background~\cite{Mao:2018jdo,Ferrer_prl,Ferrer_NPB,Ferrer_prd}. However, the dynamical generation of AMM happens only at non zero magnetic field and it is difficult to evaluate the radiative corrections to the mass and the magnetic moment independently due to single spin orientation of the Fermions in the LLL~\cite{Ferrer_NPB,Ferrer_prd}. Therefore it is unlikely that, this approach when applied to quarks will reproduce the correct value of the AMM of neutron and proton because of the absence of the strong interaction contribution. Moreover, using gauged NJL model, it was shown in Ref.~\cite{JPSingh} that the AMM of quarks can be significant in theories where mass generation occurs through dynamical chiral symmetry breaking.

Another alternative approach is to use the AMM of quarks calculated using the constituent quark model (CQM)~\cite{Halzen:1984mc,Bicudo:1998qb}, where the experimental values of the nucleon AMM are used to extract the AMM of the quarks. This procedure has already been used in Refs.~\cite{Sadooghi,Chaudhuri,Chaudhuri2} and a substantial modification in thermodynamical quantities are observed. In Ref.~\cite{Chaudhuri}, masses of scalar and neutral pseudoscalar mesons are also examined using NJL model and the Mott transition temperature is found to decrease substantially with the increase in magnetic field when the AMM of the quarks are taken into consideration. In Ref.~\cite{Snigdha2}, NJL model with non zero AMM of quarks is used to study the dilepton production rate in the presence of an arbitrary external magnetic field. In Refs.~\cite{Mei:2020jzn,Chaudhuri2}, the authors have used the Polyakov loop extended NJL (PNJL) model to study the effect of AMM of the quarks on the phase structure of magnetized quark matter.

We reiterate that, AMM of the quarks has dominant contribution from the QCD correction to the photon-quark-antiquark ($\gamma q\qbar$) vertex function. The corresponding QED correction to the vertex is subleading (as compared to the QCD correction) and thus can not explain the large value of the AMM of the proton and neutron. Moreover, due to the large QCD coupling, it is not possible to evaluate the $\gamma q\qbar$ vertex function using the perturbative QCD technique specially at low temperature. As an alternative, we use the NJL model to explicitly calculate the vertex function and extract the AMM of the quarks which reproduces the correct values of the AMM of proton and neutron.

In the current work, we have modified the two-flavor NJL Lagrangian by introducing an interaction term with a Abelian gauge field via the minimal coupling, which will be considered as small perturbation to the original field theory. Using this gauged-NJL model, we have calculated the lowest order diagram which contributes to the magnetic form factor corresponding to the effective photon-quark-antiquark ($\gamma q \qbar$) vertex at finite temperature in presence of arbitrary external magnetic field in the mean field approximation (MFA). For this, the imaginary time formalism (ITF) of finite temperature field theory and Schwinger proper time formalism are implemented in the calculation of the loop graphs; the complete (including all the Landau levels) Schwinger propagator with non-zero AMM of the quarks is considered. NJL model being non-renormalizable~\cite{Klevansky}, we have used a proper regularization scheme that correctly reproduces the well known phenomenological quantities at zero temperature and zero magnetic field such as pion-decay constant ($f_\pi$), vacuum quark condensate, vacuum pion mass ($m_\pi$) as well as the magnetic moments of proton and neutron using CQM. We then study the thermo-magnetic modification of the AMM of the quarks and found that at sufficiently high temperature both the constituent quark mass and AMM of the the quarks asymptotically vanish.

The article is organized as follows. In Sec.~\ref{sec.amm.vac}, the magnetic form factors for the effective $\gamma q \qbar$ vertex are calculated at finite temperature and magnetic field. Next in Sec.~\ref{sec.gap}, the AMM of the quarks and the constituent quark mass are extracted by solving a set of coupled gap equations. All the numerical results are presented in Sec.~\ref{sec.results} followed by a brief summary in Sec.~\ref{sec.summary}. Some of the relevant calculational details are provided in the appendix.

\section{THE PHOTON-QUARK-ANTIQUARK VERTEX FUNCTION AND THE MAGNETIC FORM FACTORS} \label{sec.amm.vac}
Let us start with the standard expression of the two-flavor gauged NJL Lagrangian
\begin{eqnarray}
\scrL_\text{NJL} = \psibar\FB{i\gamma^\mu\del_\mu - |e|\hat{Q}\gamma^\mu A_\mu -m}\psi 
+ G \SB{(\psibar\psi)^2+(\psibar i\gamma^5\bm{\tau}\psi)^2} 
\label{eq.lagrangian}
\end{eqnarray}
where, $\psi=\begin{pmatrix} \psi_\tu \\ \psi_\td \end{pmatrix}$ is the quark isospin flavor doublet with $\psi_\tu$ and $\psi_\td$ being the 
up and down quark fields respectively. In the above equation, $\hat{Q}= \begin{pmatrix} Q_\tu & 0 \\ 0 & Q_\td \end{pmatrix}$ is the charge-fraction matrix in the 
flavor space with $Q_\tu = 2/3$ and $Q_\td=-1/3$, $\bm{\tau}$ are the three Pauli isospin matrices, $|e|$ is the electric charge of a proton, $G$ is the coupling constant in the scalar channel 
for the four point contact interactions among the quark fields and $m$ is the current quark mass which is assumed to be equal for the 
up and down quarks ensuring the isospin symmetry. Throughout the paper, we have used the metric tensor with signature $g^\munu=\text{diag}(1,-1,-1,-1)$.

\begin{figure}[h]
	\begin{center}
		\includegraphics[angle=0,scale=1.0]{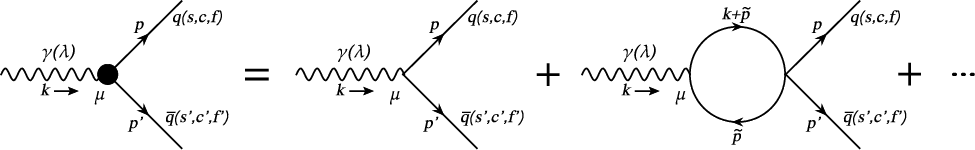}
	\end{center}
	\caption{The lowest order Feynman diagrams contributing to the $\gamma q\qbar$ vertex function. The blob represents the effective vertex $\Gamma^\mu$. All the Fermions lines denote the dressed quark propagation in a mean field. The arrows denote the direction of the momentum. }
	\label{fig.Feynman}
\end{figure}

In order to calculate the magnetic form factors and AMM of the quarks, we need to evaluate the photon-quark-antiquark ($\gamma q\qbar$) vertex function. 
For this, let us consider an initial state $\ket{I}=\ket{\gamma(k;\lambda)}$ containing a photon of momentum $k$ and polarization $\lambda$ going to a final state 
$\ket{F}=\ket{q(p;s,c,f)~\qbar(p';s',c',f')}$ containing a quark and antiquark having momenta $p$ and $p'$, spin  $s$ and $s'$, color $c$ and $c'$ and flavor $f$ and $f'$ respectively. The amplitude for the transition $\ket{I}\to\ket{F}$ is $\mathcal{S}_{FI}=\expectationvalue{F}{\hat{\mathcal{S}}}{I}$ where $\hat{\mathcal{S}}$ is the scattering matrix operator given by
\begin{eqnarray}
\hat{\mathcal{S}} = \mcT \TB{\exp\SB{i \!\int\! \FB{\scrL_e(x) + \scrL_G(x)\frac{}{}} d^4x}}
\end{eqnarray}
in which, $\mcT$ denotes the time-ordering, $\scrL_e = -|e|\psibar\hat{Q}\gamma^\mu\psi A_\mu$ and 
$\scrL_G = G \SB{(\psibar\psi)^2+(\psibar i\gamma^5\bm{\tau}\psi)^2} $. Expanding $\mathcal{S}_{FI}$ up to second order, we get after some simplifications
\begin{eqnarray}
\mathcal{S}_{FI} = i\!\!\int\!\! d^4x \expectationvalue{F}{\mcT \scrL_e(x)}{I} - \int\!\!\!\!\int \!\!d^4x d^4y \expectationvalue{F}{\mcT \scrL_e(x) \scrL_G(y)}{I} + \cdots.
\label{eq.sfi}
\end{eqnarray}
The calculations of the matrix elements are provided in Appendix~\ref{app.mat.elem} and we obtain the non-trivial contribution to the matrix elements from Eqs.~\eqref{eq.sfi.e.1} and \eqref{eq.sfi.eG.1} as
\begin{eqnarray}
\expectationvalue{F}{\mcT \scrL_e(x)}{I} &=& e^{ix\cdot(p+p'-k)}\ubar(p;s,c,f)  \FB{-|e|\hat{Q}\gamma^\mu \frac{}{}}_{c,c'}^{f,f'} v(p';s',c',f') \epsilon_\mu(k;\lambda) 
\label{eq.sfi.e}\\
 \expectationvalue{F}{\mcT \scrL_e(x) \scrL_G(y)}{I} &=& e^{iy\cdot(p+p')} e^{-ix\cdot k} \ubar(p;s,c,f)  
 \FB{-2GS(y,x)|e|\hat{Q}\gamma^\mu S(x,y) \right. \nn \\
 	&& \left. + 2G\gamma^5\tau^i S(y,x)|e|\hat{Q}\gamma^\mu S(x,y) \gamma^5\tau^i \frac{}{}}_{c,c'}^{f,f'} v(p';s',c',f') \epsilon_\mu(k;\lambda)
 \label{eq.sfi.eG}
\end{eqnarray}
where, $\epsilon_\mu(k;\lambda)$ is the polarization vector of the incoming photon, $\ubar(p;s,c,f) $ and $v(p';s',c',f') $ are respectively the color-iso-spinors representing the outgoing quark and antiquark respectively. In the above equation, $S(x,y)$ is the coordinate space Hartree-quark propagator (dressed) in vacuum given by
\begin{eqnarray}
S(x,y) = S(x-y)= \expectationvalue{\Omega}{\mcT \psi (x) \psibar (y)}{\Omega} 
\label{eq.propagator.x}
\end{eqnarray}
where $\ket{\Omega}$ denotes the NJL-vacuum corresponding to the propagation of the quarks in a Mean Field (MF). 
It is to be noted that, $S(x,y)$ is translationally invariant and is diagonal in both the color and flavor spaces. Substituting Eqs.~\eqref{eq.sfi.e} and \eqref{eq.sfi.eG} into Eq.~\eqref{eq.sfi} and performing the space-time integrals we arrive at
\begin{eqnarray}
\mathcal{S}_{FI} = (2\pi)^4\delta^4(k-p-p') \ubar(p;s,c,f) \Big(-i|e|\Gamma^\mu\Big)_{c,c'}^{f,f'} v(p';s',c',f') \epsilon_\mu(k;\lambda) + \cdots
\label{eq.sfi.1}
\end{eqnarray}
where,
\begin{eqnarray}
\Gamma^\mu(k,p) = \gamma^\mu\otimes\hat{Q}\otimes\identity_\text{Color} -2iG\!\int\!\!\!\frac{d^4\ptilde}{(2\pi)^4}\Big[S(k+\ptilde)\hat{Q}\gamma^\mu S(\ptilde)
-\gamma^5\tau^iS(k+\ptilde)\hat{Q}\gamma^\mu S(\ptilde)\gamma^5\tau^i\Big] \label{eq.vertex}
\end{eqnarray}
in which $S(p)$ is the momentum space quark Feynman propagator in vacuum given by
\begin{eqnarray}
S(p)  = i\!\!\int\!\! d^4x e^{ip.(x-y)} S(x-y) = \begin{pmatrix}
S_\tu(p) & 0 \\ 0 & S_\td(p) 
\end{pmatrix} = \frac{-(\cancel{p}+M)}{p^2-M^2+i\epsilon}\otimes\identity_\text{Flavor}\otimes\identity_\text{Color}
\label{eq.propagator}
\end{eqnarray}
with $M$ being the ``\textit{constituent quark mass}''. Eq.~\eqref{eq.sfi.1} has been represented in terms of the Feynman diagram in Fig.~\ref{fig.Feynman}.

In order to extract the magnetic form factors and AMM of the quarks from the effective vertex $\Gamma^\mu(k,p)$, we decompose it as
\begin{eqnarray}
\Gamma^\mu(k,p) = \Big[F_1(k,p)~\hat{Q}\otimes\gamma^\mu + F_2(k,p)~\hat{Q}~i \frac{\sigma^\munu}{2M}k_\nu \Big] \otimes\identity_\text{Color} + \cdots
\end{eqnarray}
where $F_1$ and $F_2$ are respectively the electric and magnetic form factors. The quantity $F_2$ can be easily extracted from $\Gamma^\mu(k,p)$ as
\begin{eqnarray}
F_2(k,p) = -i\frac{M}{6N_ck^2}\hat{Q}^{-1}\Tr_\text{d,c}\FB{k_\nu\sigma^\munu\Gamma_\mu} \label{eq.F2}
\end{eqnarray}
where, $N_c$ is the number of color and the subscripts `d' and `c' in trace correspond to the traces taken over the Dirac and color spaces respectively. Substituting Eq.~\eqref{eq.vertex} into \eqref{eq.F2}, we get after some simplifications
\begin{eqnarray}
F_2(k) = \begin{pmatrix}
F_2^\tu & 0 \\ 0 & F_2^\td
\end{pmatrix} = 
-\frac{MG}{3N_ck^2}\int\frac{d^4\ptilde}{(2\pi)^4}\begin{pmatrix}
\Tr_\text{d,c}[k_\nu\sigma^\munu S_\td(k+\ptilde)\gamma_\mu S_\td(\ptilde)] & 0 \\
0 & 4\Tr_\text{d,c}[k_\nu\sigma^\munu S_\tu(k+\ptilde)\gamma_\mu S_\tu(\ptilde)]
\end{pmatrix}.
\label{eq.F2.1}
\end{eqnarray}
Substituting $S(p)$ from Eq.~\eqref{eq.propagator} into the above equation leads to the following simplified expression of the magnetic form factors corresponding to the up and down quarks
\begin{eqnarray}
F_2^\tu(k) = \frac{1}{4}F_2^\td(k) = -4GM^2 i \!\! \int \!\!\! \frac{d^4\ptilde}{(2\pi)^4} \frac{1}{\{(\ptilde+k)^2-M^2+i\epsilon\}(\ptilde^2-M^2+i\epsilon)}.
\end{eqnarray}

For the incorporation finite temperature in the calculation of the form factors, we employ the Imaginary Time Formalism (ITF) of finite temperature field theory in which the continuous energies are replaced by discrete Matsubara modes as $\ptilde^0\to(2r+1)i\pi T$ where $T$ is the temperature and $ r \in \mathbb{Z} $. 
Thus the magnetic form factors at finite temperature become
\begin{eqnarray}
F_2^\tu(k;T) = \frac{1}{4}F_2^\td(k;T) = 4GM^2T \!\!\int\!\!\! \frac{d^3\ptilde}{(2\pi)^3}\!\sum_{r=-\infty}^{\infty} \frac{1}{\{(\ptilde+k)^2-M^2+i\epsilon\}(\ptilde^2-M^2+i\epsilon)}\Bigg|_{\ptilde^0=(2r+1)i\pi T}. 
\end{eqnarray}
Performing the sum over Matsubara frequencies, the above equation leads to
\begin{eqnarray}
F_2^\tu(k;T) = \frac{1}{4}F_2^\td(k;T) = -2GM^2 \!\!\int\!\!\! \frac{d^3\ptilde}{(2\pi)^3} \Bigg[
\frac{(\wptd+\wktd)}{\wptd\wktd}\frac{1}{k_0^2-(\wptd+\wktd)^2}
- \frac{f(\wptd)/\wptd}{(k_0+\wptd)^2-\wktd^2} - \frac{f(\wptd)/\wptd}{(k_0-\wptd)^2-\wktd^2} \nn \\
- \frac{f(k_0+\wktd)/\wktd}{(k_0+\wktd)^2-\wptd^2} - \frac{f(-k_0+\wktd)/\wktd}{(k_0-\wktd)^2-\wptd^2} \Bigg]
\label{eq.F2.T}
\end{eqnarray}
where, $\wptd=\sqrt{\ptdvec^2+M^2}$, $\wktd=\sqrt{\ktdvec^2+M^2}=\sqrt{(\bm{k}+\ptdvec)^2+M^2}$ and $f(x)=\TB{e^{x/T}+1}^{-1}$ is the Fermi-Dirac thermal distribution function of the quarks.

Let us now consider a constant external magnetic field $\bm{B}=B\hat{z}$ along the positive $z$-direction. 
The incorporation of such background classical field in the evaluation of the vertex function can be done using the Schwinger proper-time formalism in which the quark propagator of Eq.~\eqref{eq.propagator.x} modifies to
\begin{eqnarray}
S_B(x,y) = \Phi(x,y)S_B(x-y)= \expectationvalue{\Omega_B}{\mcT \psi (x) \psibar (y)}{\Omega_B} 
\label{eq.propagator.xB}
\end{eqnarray}
where, $\ket{\Omega_B}$ denotes the magnetized NJL-vacuum in MF approximation. 
It is to be noted that, $S_B(x,y)$ is not translationally invariant due to the presence of the phase factor $\Phi(x,y)$ but is diagonal in both the color and flavor spaces. However, the substitution of Eq.~\eqref{eq.propagator.xB} into Eqs.~\eqref{eq.sfi.e} and \eqref{eq.sfi.eG} leads to the 
cancellation of the two phase factors coming from the two propagators since the phase factor satisfies $\Phi(y,x)\Phi(x,y)=1$. Therefore, we can work with the  translationally invariant piece $S_B(x-y)$ of the Schwinger propagator for the calculation of the vertex function and analogously obtain
\begin{eqnarray}
\Gamma^\mu(k,p) = \gamma^\mu\otimes\hat{Q}\otimes\identity_\text{Color} -2iG \!\!\int\!\!\!\frac{d^4\ptilde}{(2\pi)^4}\Big[S_B(k+\ptilde)\hat{Q}\gamma^\mu S_B(\ptilde)
-\gamma^5\tau^iS_B(k+\ptilde)\hat{Q}\gamma^\mu S_B(\ptilde)\gamma^5\tau^i\Big] \label{eq.vertex.B}
\end{eqnarray}
in which $S_B(p)$ is the momentum space quark Schwinger propagator in vacuum given by
\begin{eqnarray}
S_B(p)  = i\!\!\int\!\!\! d^4x e^{ip.(x-y)} S_B(x-y) = \begin{pmatrix}
S^B_\tu(p) & 0 \\ 0 & S^B_\td(p) 
\end{pmatrix}
\label{eq.propagator.B}
\end{eqnarray}
where each of the diagonal flavor component becomes sum over discrete Landau levels and spin as
\begin{eqnarray}
S^B_f(p) = \sum_{s\in\{\pm1\}}^{} \sum_{n=0}^{\infty} \frac{-\scrD_{nfs}(p)}{\ppll^2-M_{nfs}^2+i\epsilon}\otimes\identity_\text{Color}
~~;~f\in\{\tu,\td\}.
\label{eq.propagator.1}
\end{eqnarray}
In the above equation, $M_{nfs}=\MB{M_{nf}-s\kappa_fQ_fB}$ where $M_{nf}=\sqrt{M^2+2n|Q_feB|}$ with $\kappa_f$ being the AMM of quark flavor $f$. 
With respect to the direction of the external magnetic field, we have decomposed $p=(\ppll+\pper)$ where $\ppll^\mu = \gpll^\munu p_\nu$ and 
$\pper^\mu = \gper^\munu p_\nu$ with $\gpll^\munu=\text{diag}(1,0,0,-1)$ and $\gper^\munu=\text{diag}(0,-1,-1,0)$. 
The quantity $\scrD_{nfs}(p)$ in the above equation contains the Dirac structure of the propagator and its explicit form is 
\begin{eqnarray}
\scrD_{nfs}(p) = (-1)^ne^{-\alpha_p^f}\frac{1}{2M_{nf}}(1-\delta_n^0\delta_s^{-1})
\Big[(M_{nf}+sM)(\cancel{p}_\parallel-\kappa_fQ_fB+sM_{nf}) \FB{\identity+\text{sign}(Q_f)i\gamma^1\gamma^2}L_n(2\alpha_p^f) \nn \\
-(M_{nf}-sM)(\cancel{p}_\parallel-\kappa_fQ_fB-sM_{nf})\FB{\identity-\text{sign}(Q_f)i\gamma^1\gamma^2}L_{n-1}(2\alpha_p^f) \nn \\
-4s \FB{ \cancel{p}_\parallel - \text{sign}(Q_f)i\gamma^1\gamma^2 (\kappa_fQ_fB-sM_{nf})} \text{sign}(Q_f)i\gamma^1\gamma^2 \cancel{p}_\perp
L_{n-1}^1(2\alpha_p^f)   \Big] \label{eq.Dnfs}
\end{eqnarray}
where, $\alpha_p^f=-p_\perp^2/|Q_feB|$ and $L_n^a(z)$ denotes the associate Laguerre polynomials with convention $L_{-1}^a(z)=0$. This convention along with the presence of the factor $(1-\delta_n^0\delta_s^{-1})$ in Eq.~\eqref{eq.Dnfs} ensure that the Lowest Landau Level (LLL) is spin non-degenerate. Substituting Eq.~\eqref{eq.vertex.B} into \eqref{eq.F2}, we obtain the magnetic form factor analogous to Eq.~\eqref{eq.F2.1} which are now functions of external magnetic field $B$ and AMM of the quark as
\begin{eqnarray}
	F_2(k;B,\kappa_\tu,\kappa_\td) = \begin{pmatrix}
		F_2^\tu(k;B,\kappa_\td) & 0 \\ 0 & F_2^\td(k;B,\kappa_\tu)
	\end{pmatrix} = 
	-\frac{MG}{3N_ck^2}\begin{pmatrix}
		\text{I}_\td(k;B,\kappa_\td) & 0 \\
		0 & 4\text{I}_\tu(k;B,\kappa_\tu)
	\end{pmatrix}
	\label{eq.F2.2}
\end{eqnarray}
where, 
\begin{eqnarray}
\text{I}_f(k;B,\kappa_f) = \!\!
\int\!\!\!\frac{d^4\ptilde}{(2\pi)^4}\Tr_\text{d,c}[k_\nu\sigma^\munu S^B_f(k+\ptilde)\gamma_\mu S^B_f(\ptilde)] ~~;~f\in\{\tu,\td\}.
\end{eqnarray}
The incorporation of the effect of finite temperature can now be done using ITF in which we again replace continuous energies by discrete Matsubara modes as $\ptilde^0\to(2r+1)i\pi T$. Thus Eqs.~\eqref{eq.F2.2} modifies to
\begin{eqnarray}
F_2^\tu(k;B,\kappa_\td,T) =  -\frac{MG}{3N_ck^2}\text{I}_\td(k;B,\kappa_\td,T) \label{eq.F2.u}\\
F_2^\td(k;B,\kappa_\tu,T) =  -\frac{MG}{3N_ck^2}\text{I}_\tu(k;B,\kappa_\tu,T) \label{eq.F2.d}
\end{eqnarray}
where, 
\begin{eqnarray}
\text{I}_f(k;B,\kappa_f,T) = 
\!\!\int\!\!\!\frac{d^3\ptilde}{(2\pi)^3} ~iT\sum_{r=-\infty}^{\infty} \Tr_\text{d,c}[k_\nu\sigma^\munu S^B_f(k+\ptilde)\gamma^\mu S^B_f(\ptilde)]
\Bigg|_{\ptilde^0=(2r+1)i\pi T} ~~;~f\in\{\tu,\td\}.
\label{eq.If}
\end{eqnarray}
For simplicity in analytic calculation, we will take $\kper=0$. Substituting Eq.~\eqref{eq.propagator.1} into the above equation, we get after a long but straightforward calculation the quantity $\text{I}_f(\kpll,\kper=0;B,\kappa_f,T) $ as 
\begin{eqnarray}
\text{I}_f(\kpll;B,\kappa_f,T) &=&  \frac{i}{4\pi}N_c\sum_{s_k\in\{\pm1\}}^{} \sum_{s_p\in\{\pm1\}}^{} \sum_{l=0}^{\infty} \sum_{n=0}^{\infty}
\frac{|Q_feB|s_ks_p}{M_{lfs_k}M_{nfs_p}}(1-\delta_l^0\delta_{s_k}^{-1})(1-\delta_n^0\delta_{s_p}^{-1}) \nn \\ && \hspace{-2cm} 
\times \Big[\big\{(s_kM_{lf}-s_pM_{nf})\text{I}_1 + (s_kM_{lf}-\kappa_fQ_fB)\kpll^2 \text{I}_2\big\} \big\{ (M+s_kM_{lf})(M+s_pM_{nf})\delta_l^n - (M-s_kM_{lf})(M-s_pM_{nf})\delta_{l-1}^{n-1} \big\} \nn \\ && 
-2\big\{(s_kM_{lf}+s_pM_{nf}-2\kappa_fQ_fB)\text{I}_1 + (s_kM_{lf}-\kappa_fQ_fB)\kpll^2 \text{I}_2\big\} \nn \\ &&
\times \big\{ (M+s_kM_{lf})(M-s_pM_{nf})\delta_l^{n-1} - (M-s_kM_{lf})(M+s_pM_{nf})\delta_{l-1}^{n} \big\} \Big]
\label{eq.If.1}
\end{eqnarray}
where a Kronecker delta with negative index is considered to be zero (i.e. $\delta_{-1}^{-1}=0$) and $\text{I}_1$ and $\text{I}_2$ are given by 
\begin{eqnarray}
\text{I}_2 &=& \frac{1}{2}\int_{-\infty}^{\infty}\!\frac{d\ptd_z}{(2\pi)}\Bigg[
\frac{(\wl+\wn)}{\wl\wn}\frac{1}{k_0^2-(\wl+\wn)^2} 
- \frac{f(\wl)/\wl}{(k_0+\wl)^2-\wn^2} \nn \\ 
&& - \frac{f(\wl)/\wl}{(k_0-\wl)^2-\wn^2} 
- \frac{f(k_0+\wn)/\wn}{(k_0+\wn)^2-\wl^2} - \frac{f(-k_0+\wn)/\wn}{(k_0-\wn)^2-\wl^2} \Bigg], \label{eq.I2}\\
\text{I}_1 &=& \frac{1}{4}\int_{-\infty}^{\infty}\!\frac{d\ptd_z}{(2\pi)}\frac{1}{\wl\wn}\Bigg[
(\wn-\wl)- 2f(\wl)\wn  \nn \\ 
&& + \big\{f(k_0+\wn)+f(-k_0+\wn)\big\}\wl \Bigg] -\frac{1}{2}\FB{\kpll^2+M_{lfs_k}^2-M_{nfs_p}^2}\text{I}_2 \label{eq.I1}
\end{eqnarray}
in which, $\wl=\sqrt{\ptd_z^2+M_{lfs_k}^2}$ and $\wn=\sqrt{(\ptd_z+k_z)^2+M_{nfs_p}^2}$.
%

\section{THE COUPLED GAP EQUATIONS AND AMM OF THE QUARKS} \label{sec.gap}
The AMM of the quarks namely $\kappa_\tu$ and $\kappa_\td$ are related to the magnetic form factors (calculated in the previous section) by the relation
\begin{eqnarray}
\kappa_f = \frac{1}{2M} F_2^f(k\to0)~~;~f\in\{\tu,\td\}.
\label{eq.kappa}
\end{eqnarray}
At $B=0$, we can therefore calculate $\kappa_f$ by substituting Eq.~\eqref{eq.F2.T} into Eq.~\eqref{eq.kappa} provided we know the 
constituent quark mass $M=M(T)$ which is obtained by solving the gap equation in absence of external magnetic field~\cite{Klevansky},
\begin{eqnarray}
M(T) = m + 2G N_c\sum_{f\in\{\tu,\td\}} \int\!\!\frac{d^3\ptd}{(2\pi)^3}\frac{2M}{\wptd}\TB{1-2f(\wptd)}.
\label{eq.gap}
\end{eqnarray}

The situation becomes lot more complicated at $B\ne0$ where the AMM of the quarks are to be obtained from 
Eqs.~\eqref{eq.F2.u}, \eqref{eq.F2.d} and \eqref{eq.kappa} as
\begin{eqnarray}
\kappa_\tu &=& -\frac{G}{6N_ck^2}\text{I}_\td(k\to0;B,\kappa_\td,T,M), \label{eq.ku}\\
\kappa_\td &=& -\frac{G}{6N_ck^2}\text{I}_\tu(k\to0;B,\kappa_\tu,T,M) \label{eq.kd}
\end{eqnarray}
along with the constituent quark mass $M=M(T,B,\kappa_\tu,\kappa_\td)$ satisfying the following gap equation
\begin{eqnarray}
M(T,B,\kappa_\tu,\kappa_\td) = m + 2G N_c \!\!\! \sum_{f\in\{\tu,\td\}} \!\!\!\!\frac{|Q_feB|}{2\pi}\sum_{l=0}^{\infty}\sum_{s_k\in\{\pm1\}}(1-\delta_l^0\delta_{s_k}^{-1})
\FB{1-\frac{s_k\kappa_fQ_fB}{M_{lf}}} \int_{0}^{\infty}\!\frac{d\ptd_z}{2\pi}\frac{2M}{\wl}\TB{1-2f(\wl)}.
\label{eq.gap.B}
\end{eqnarray}
Therefore Eqs.~\eqref{eq.ku}, \eqref{eq.kd} and \eqref{eq.gap.B} constitute the set of three non-linear coupled equations for the three unknown quantities $\kappa_\tu$, $\kappa_\td$ and $M$ and they have to be simultaneously solved. We name Eqs.~\eqref{eq.ku}-\eqref{eq.gap.B} as the ``\textit{coupled gap equations}''.

Once the AMM of the quarks are known, few other physical quantities can be easily calculated. 
For example, the magnetic moments of the quarks are given by
\begin{eqnarray}
\mu_f = Q_f(1+2M\kappa_f)\frac{M_\text{N}}{M} \mu_\text{N} ~~;~f\in\{\tu,\td\}
\end{eqnarray}
where $M_\text{N}=938$ MeV is the nucleon mass and $\mu_\text{N}=|e|/2M_\text{N}$ is the nuclear magneton. 
Now using the constituent quark model, the magnetic moments of proton and neutron come out to be~\cite{Halzen:1984mc,Schwartz}
\begin{eqnarray}
\mu_\text{proton} &=& \frac{1}{3}\FB{4\mu_\text{u} - \mu_\text{d}}, \\
\mu_\text{neutron} &=& \frac{1}{3}\FB{4\mu_\text{d} - \mu_\text{u}}. 
\end{eqnarray}

We conclude this section by mentioning the regularization procedure used in this work. First, we note that the temperature independent parts in Eqs.~\eqref{eq.F2.T}, \eqref{eq.I2}, \eqref{eq.I1}, \eqref{eq.gap} and \eqref{eq.gap.B} are Ultra Violet (UV) divergent. The NJL model being non-renormalizable requires proper regularization procedure. 
In this work, we have used two different regularization methods namely (i) sharp cutoff and (ii) smooth cutoff. 
In sharp cutoff scheme, we use a three-momentum cutoff $\Lambda$ to regulate the UV divergences so that, at zero magnetic field,
\begin{eqnarray}
\int\!\! d^3k f(\vec{k}) \to \int\!\! d^3k f(\vec{k}) \Theta\FB{\Lambda-|\vec{k}|}
\end{eqnarray}
whereas for non-zero magnetic field, we use
\begin{eqnarray}
\sum_{l=0}^{\infty}\int_{-\infty}^{\infty} dk_z f(k_z,l) \to  \sum_{l=0}^{\infty}\int_{-\infty}^{\infty} dk_z f(k_z,l) \Theta\FB{\Lambda - \sqrt{k_z^2+2l|eB|}}.
\end{eqnarray}
On the other hand, in smooth cutoff scheme, we use the following regularization prescription~\cite{Fukushima:2010fe} at $B=0$:
\begin{eqnarray}
\int\!\! d^3k f(\vec{k}) \to \int\!\! d^3k f(\vec{k}) \sqrt{\frac{\Lambda^{20}}{\Lambda^{20}+|\vec{k}|^{20}}}
\end{eqnarray}
and for non-zero magnetic field,
\begin{eqnarray}
\sum_{l=0}^{\infty}\int_{-\infty}^{\infty} dk_z f(k_z,l) \to  \sum_{l=0}^{\infty}\int_{-\infty}^{\infty} dk_z f(k_z,l) \sqrt{\frac{\Lambda^{20}}{\Lambda^{20}+\FB{k_z^2+2l|eB|}^{10}}}.
\end{eqnarray}
%
%
\begin{table}[h]
	\begin{center}
		\begin{tabular}{cccc}
			\hline
			$\Lambda_\text{Sharp}$ & ~~~~~~~~~~~~~ $\Lambda_\text{Smooth}$ & ~~~~~~~~~~~~~ $G$ ~~~~~~~~~~~~~ & ~~~~~~~~~~~ $m$ ~~~~~~~~~~~ \\
			\hline 	\hline
			623.95 MeV & ~~~~~~~~~~~~568.69 MeV &~~~~~ 5.844 GeV$^{-2}$ & 5.6 MeV \\ 
			\hline	
		\end{tabular}
	\end{center}
	\caption{ Choice of the different parameters used in this work. }
	\label{table1}
\end{table}
\begin{figure}[h]
	\begin{center}
		\includegraphics[angle=-90,scale=0.35]{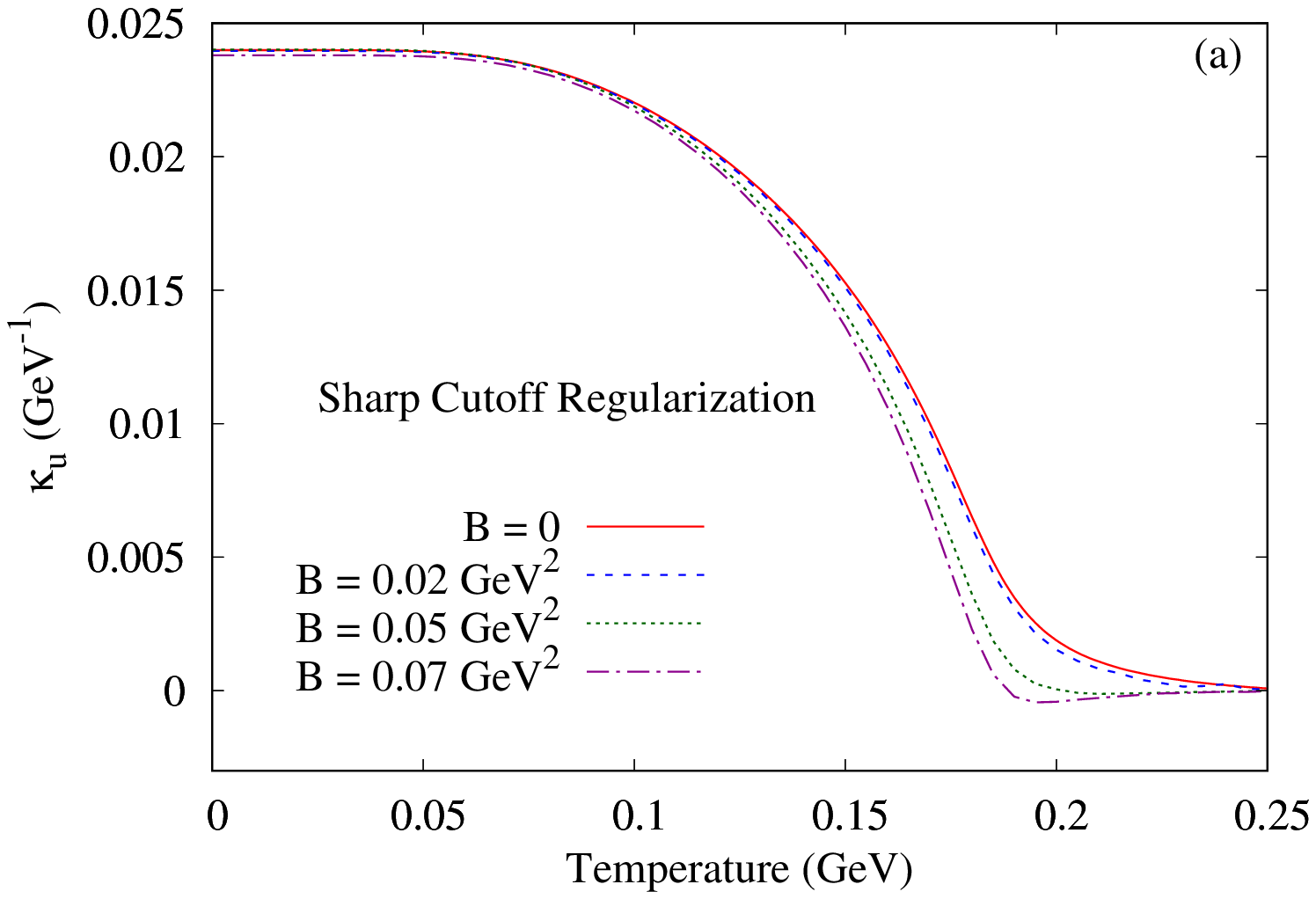} \includegraphics[angle=-90,scale=0.35]{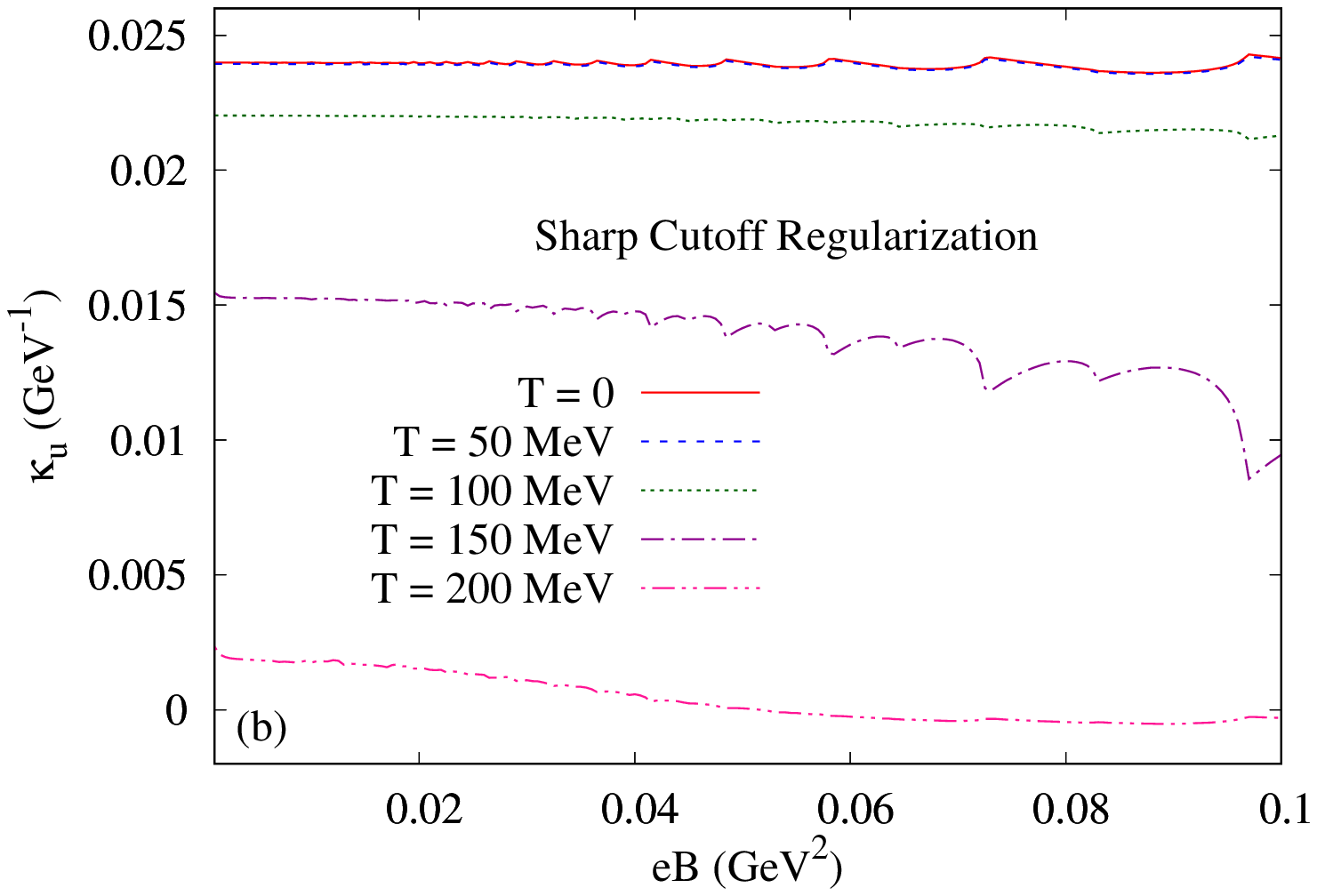} 
		\includegraphics[angle=-90,scale=0.35]{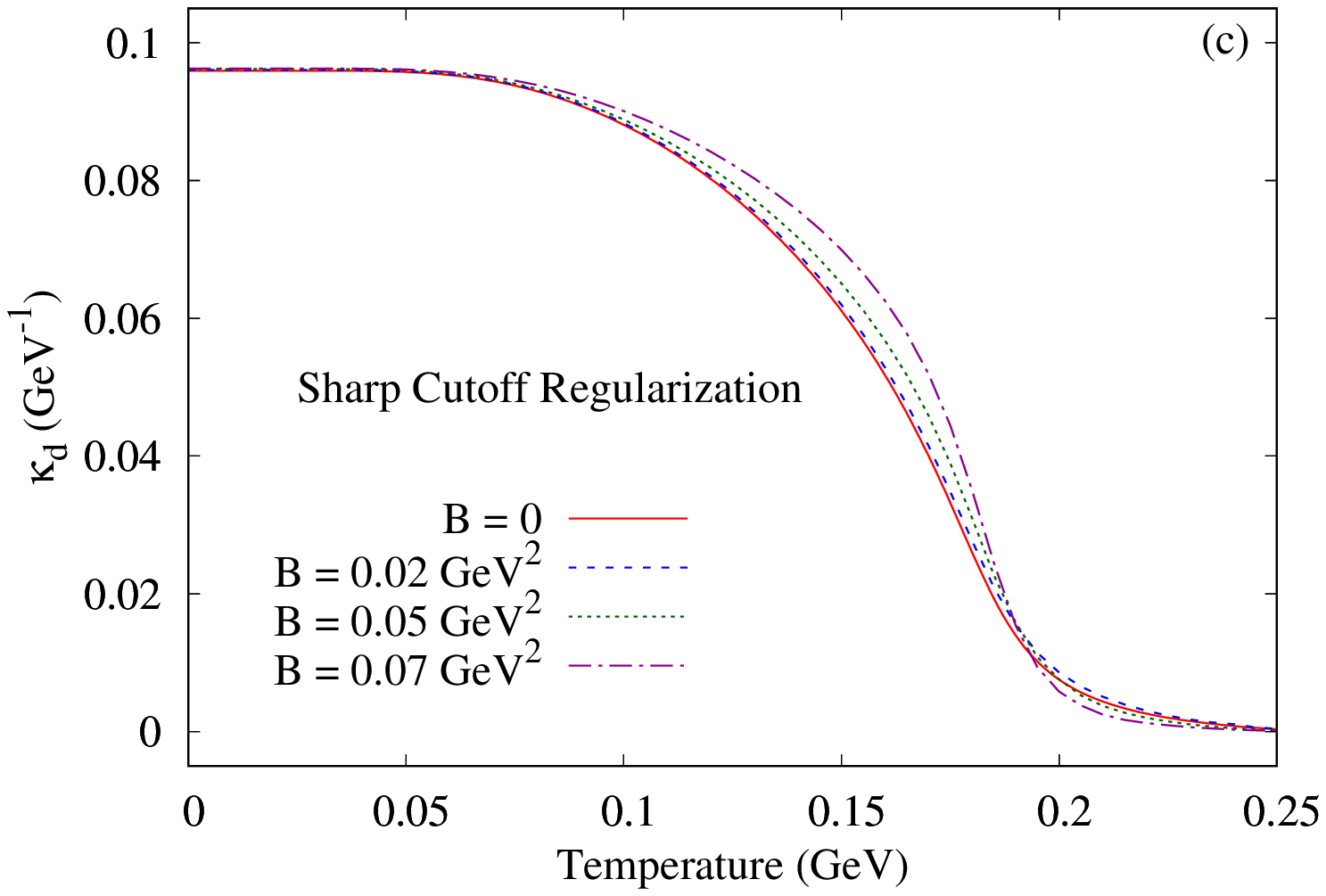} \includegraphics[angle=-90,scale=0.35]{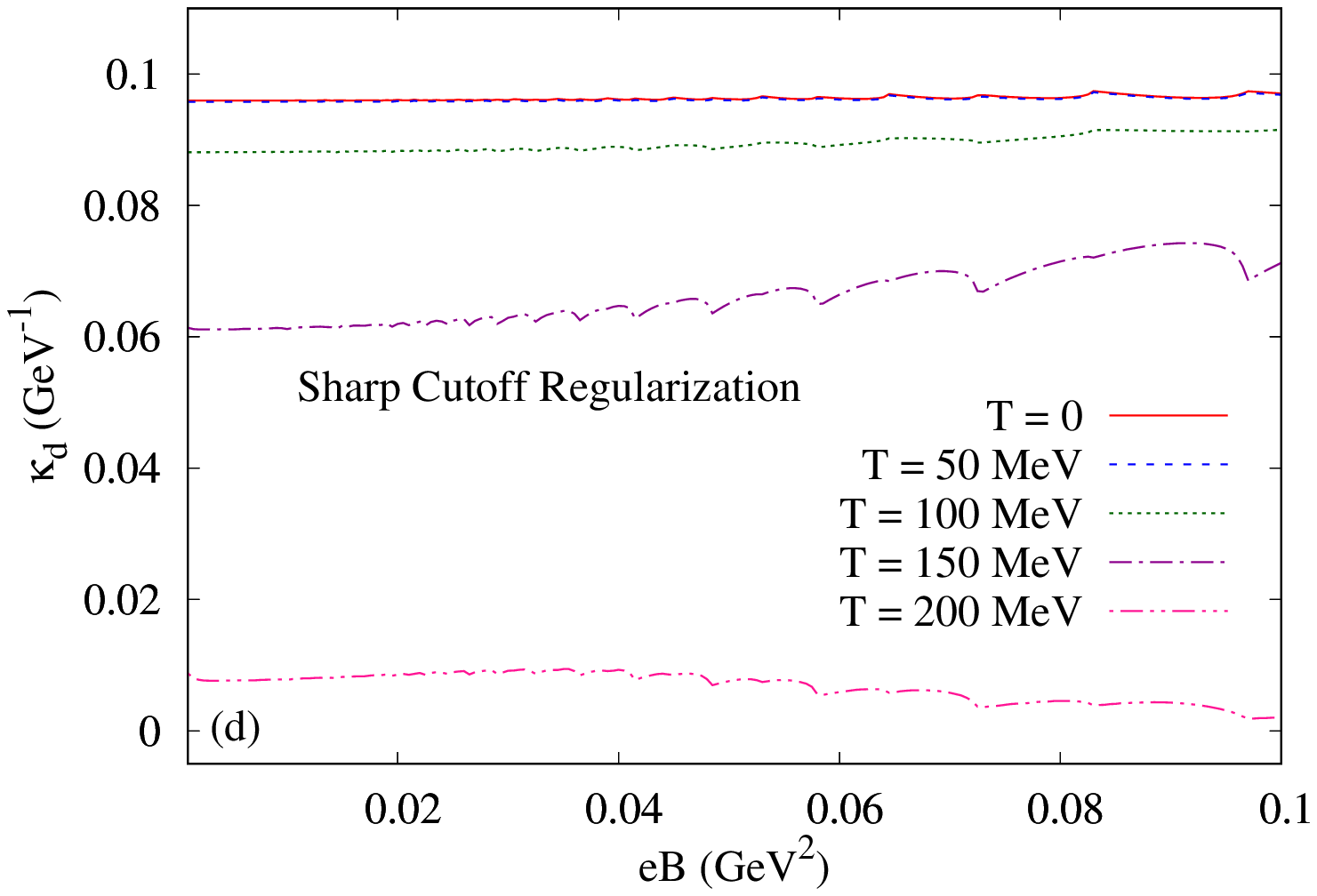}
		\includegraphics[angle=-90,scale=0.35]{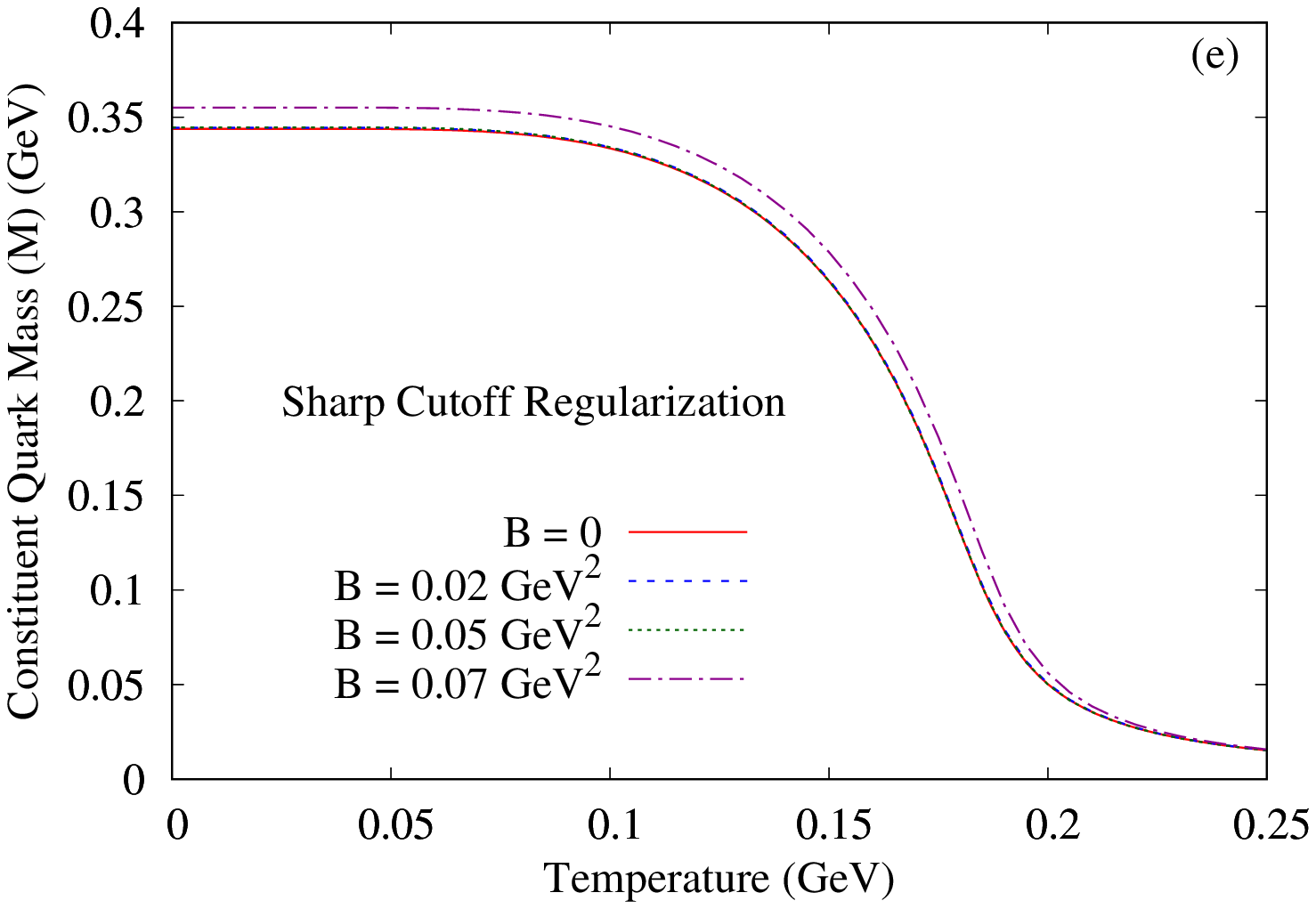} \includegraphics[angle=-90,scale=0.35]{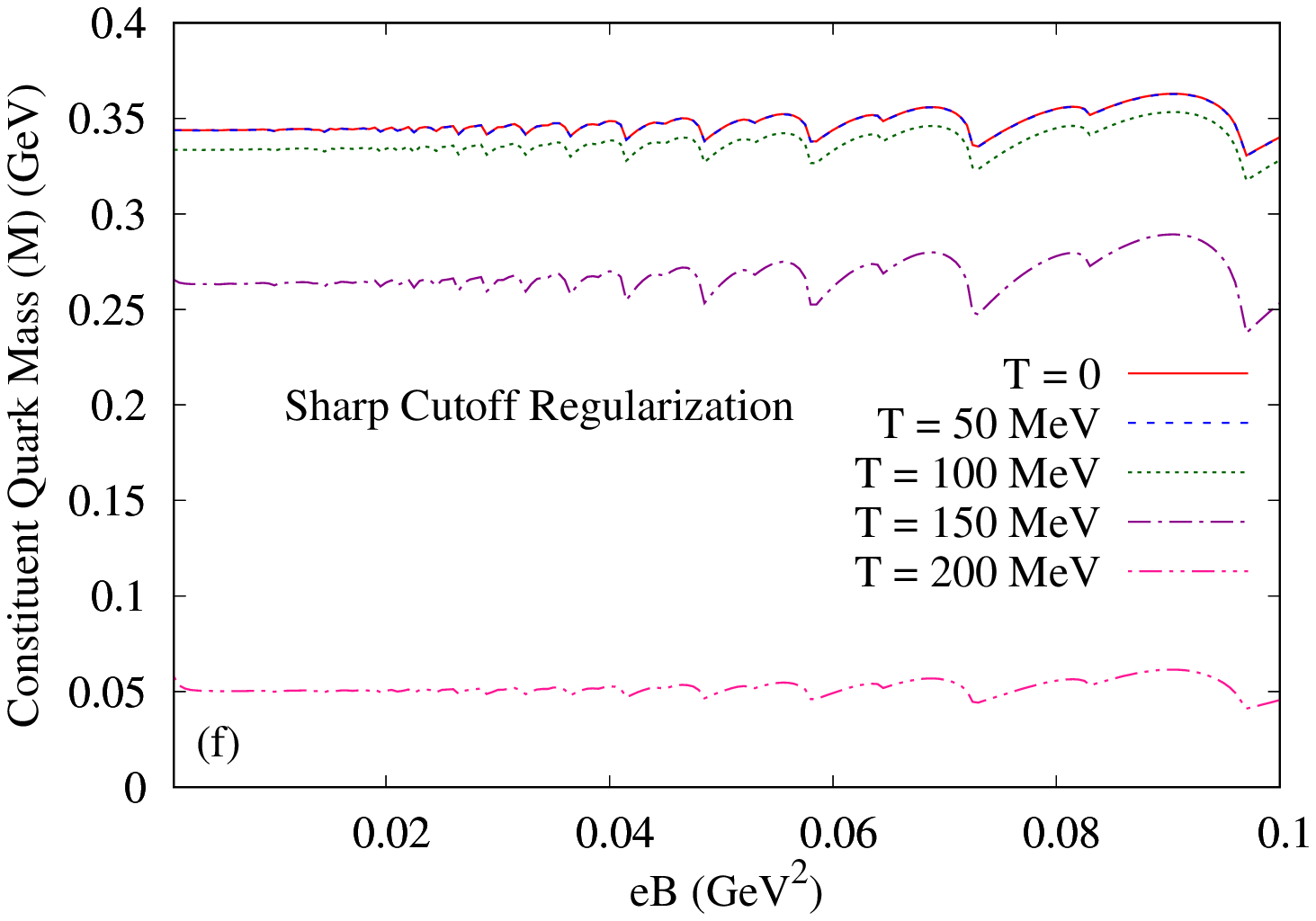}
	\end{center}
	\caption{The variation of (a) $\kappa_\text{u}$, (c) $\kappa_\text{d}$ and (e) $M$ as a function of temperature using the sharp cutoff regularization scheme. The variation of (b) $\kappa_\text{u}$, (d) $\kappa_\text{d}$ and (f) $M$ as a function of magnetic field using the sharp cutoff regularization scheme.}
	\label{fig.1}
\end{figure}
\begin{figure}[h]
	\begin{center}
		\includegraphics[angle=-90,scale=0.35]{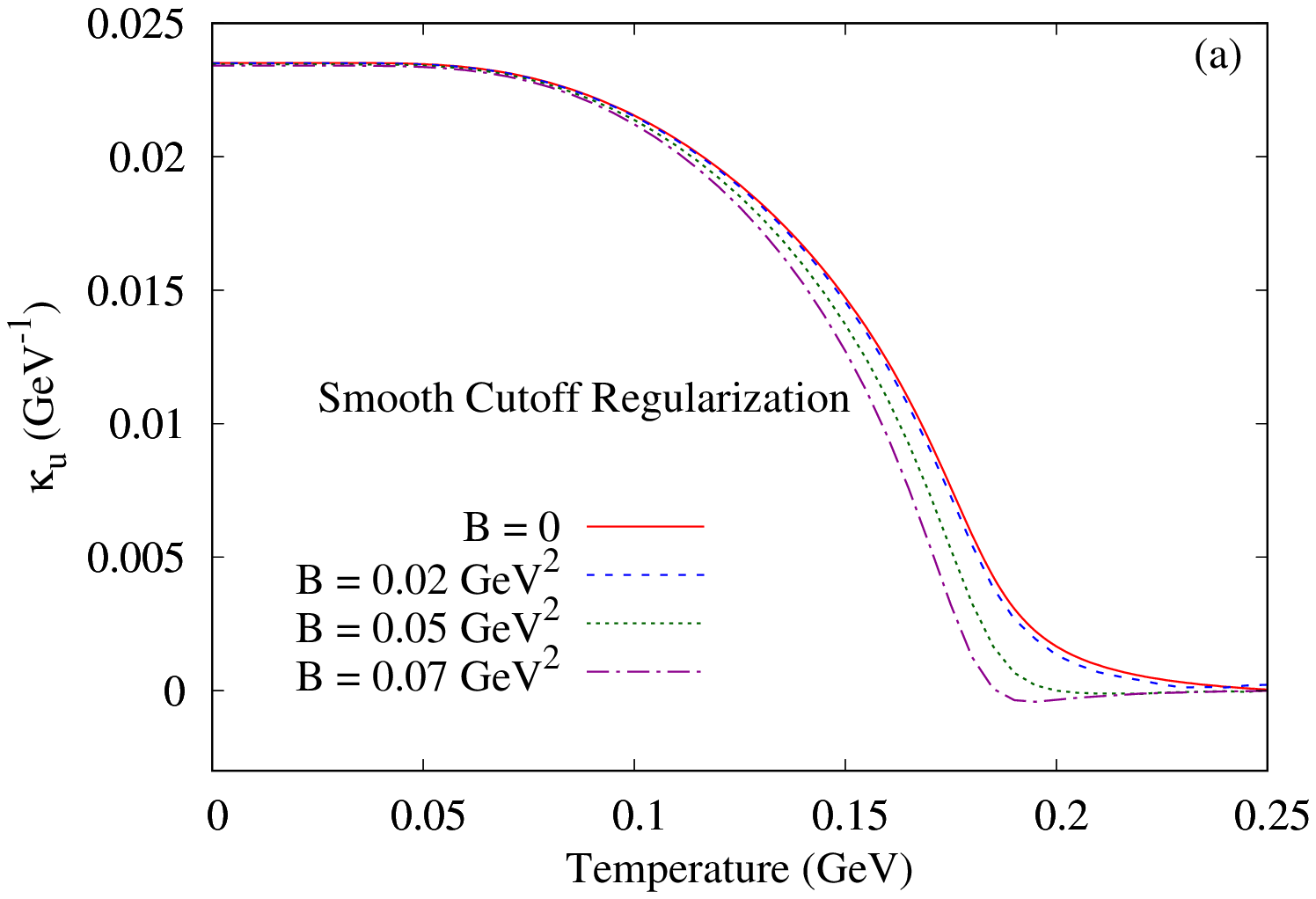} \includegraphics[angle=-90,scale=0.35]{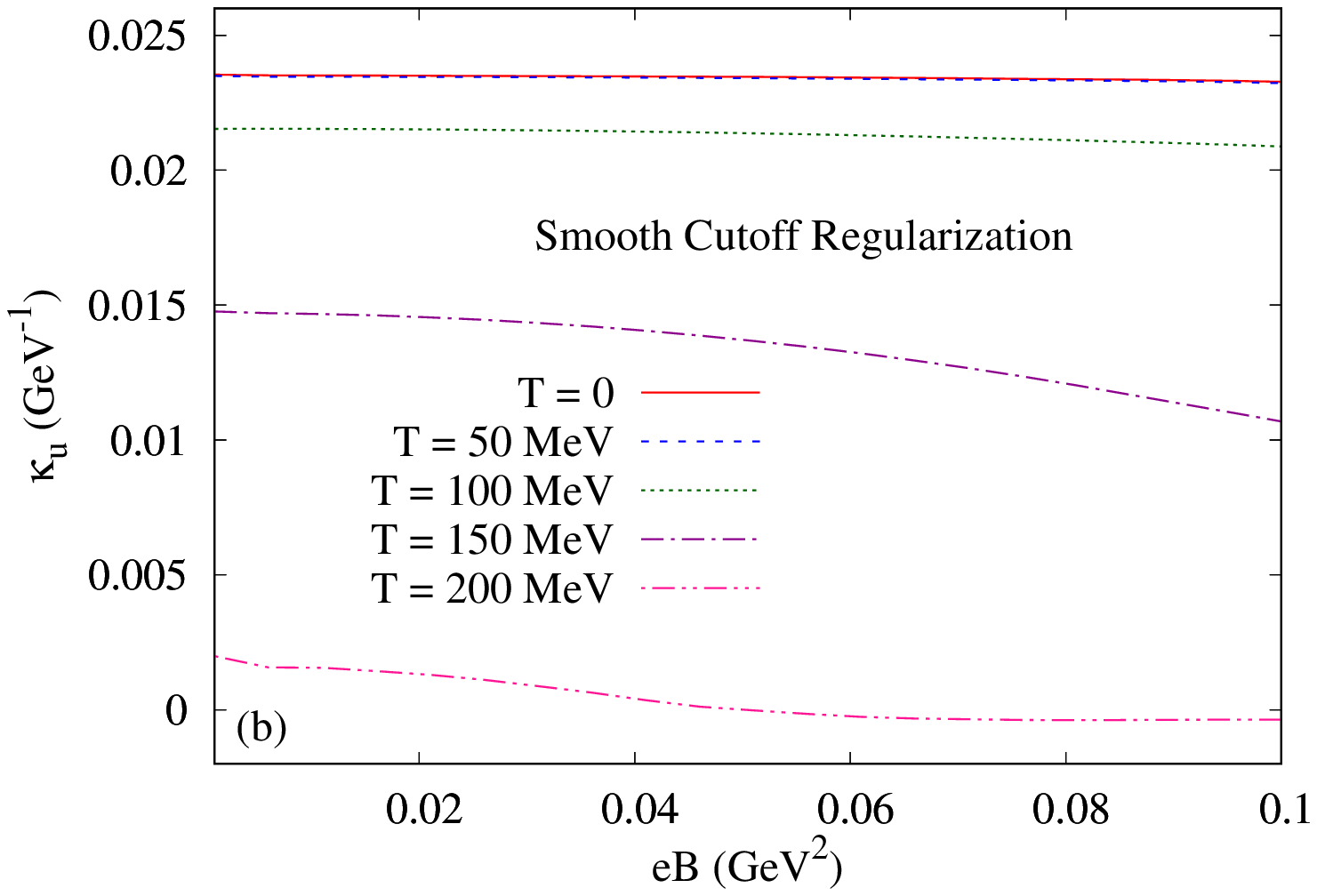} 
		\includegraphics[angle=-90,scale=0.35]{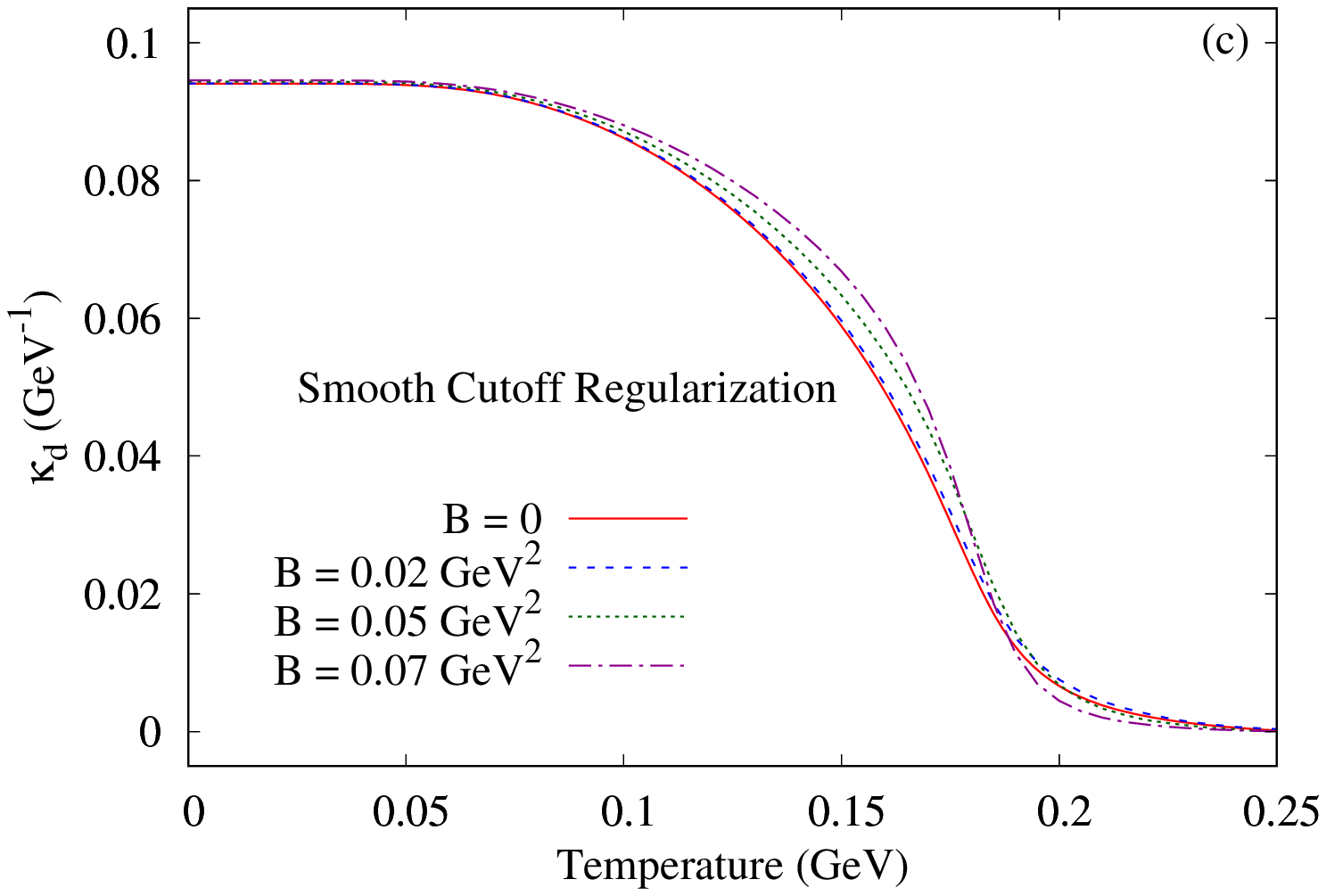} \includegraphics[angle=-90,scale=0.35]{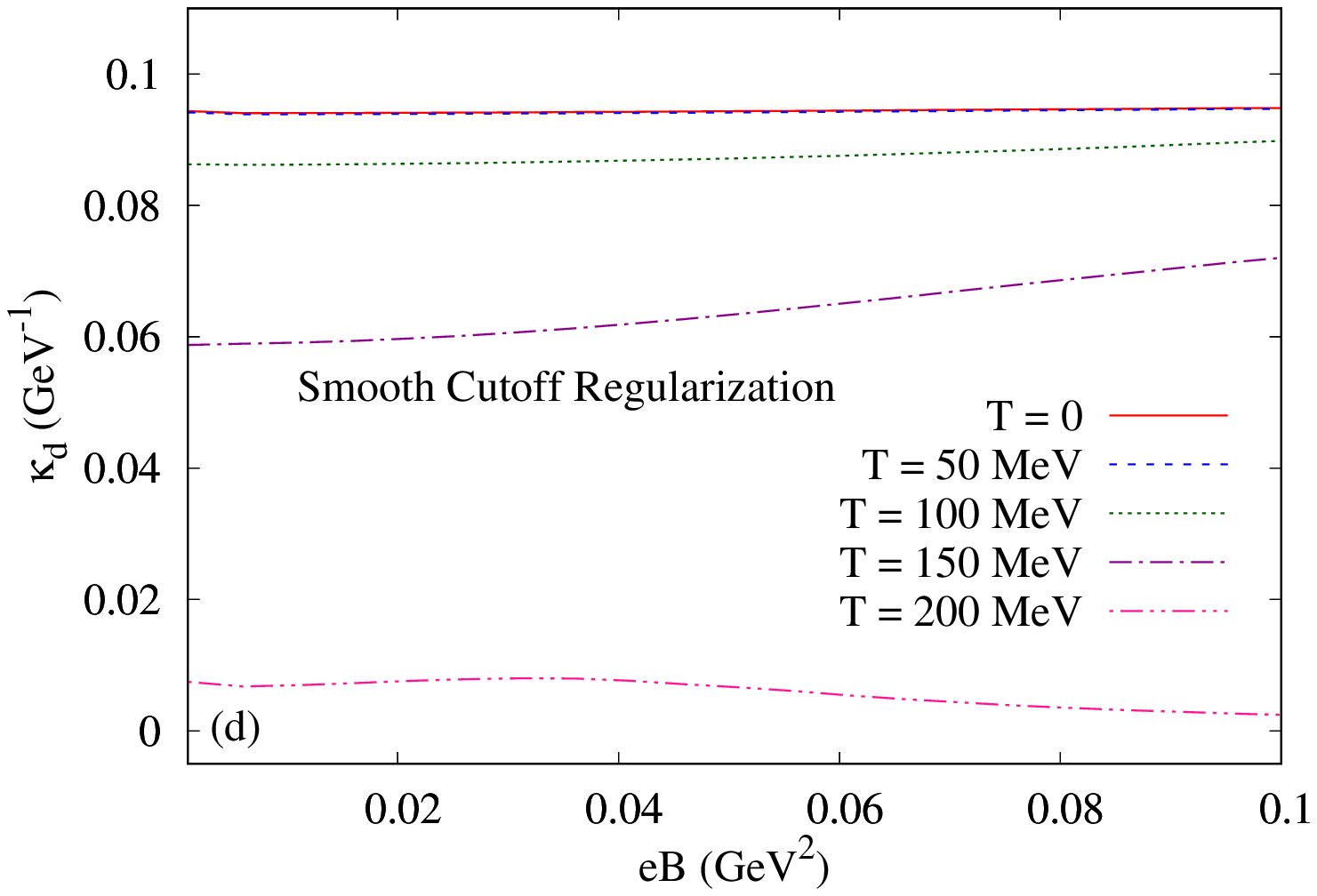}
		\includegraphics[angle=-90,scale=0.35]{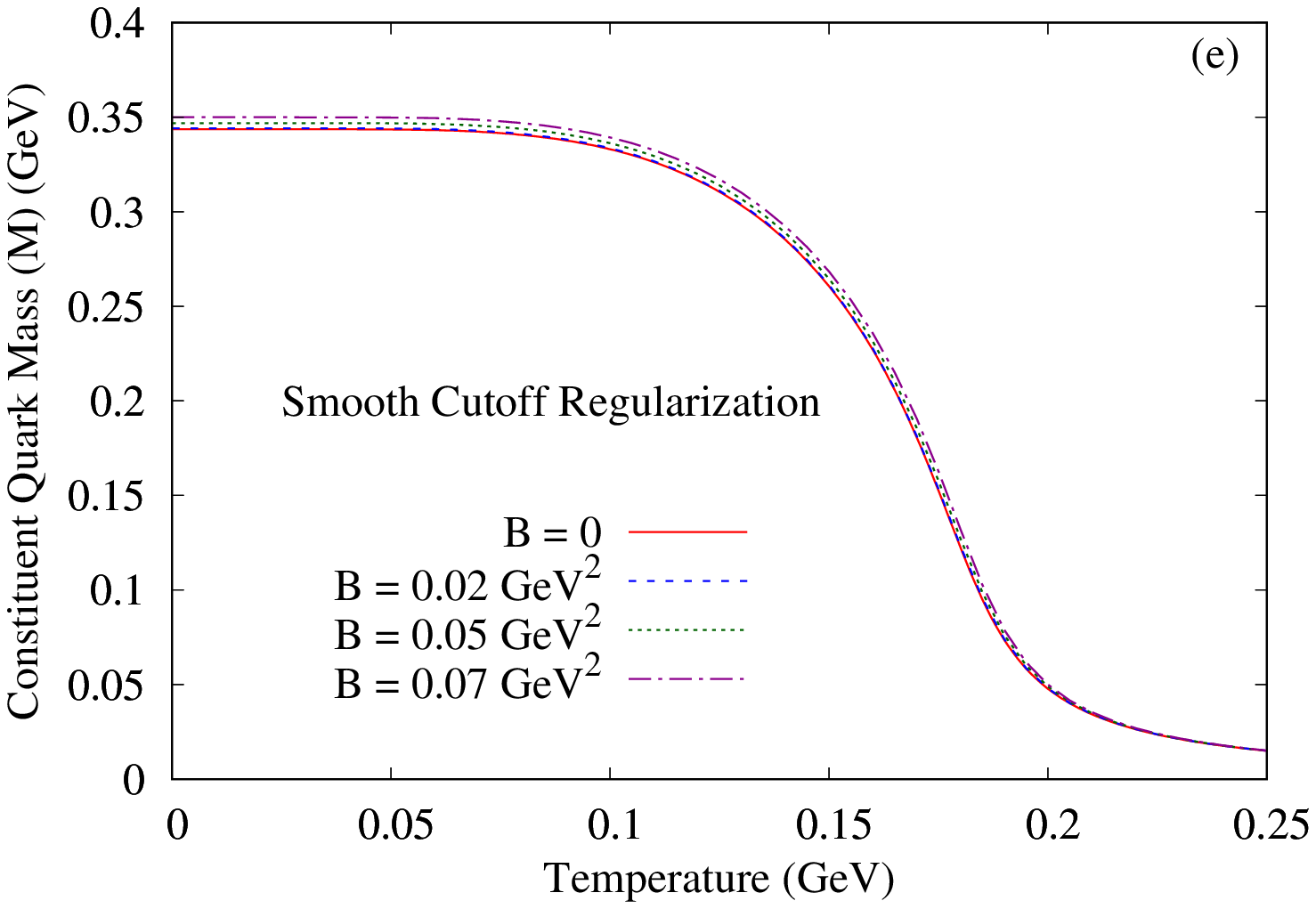} \includegraphics[angle=-90,scale=0.35]{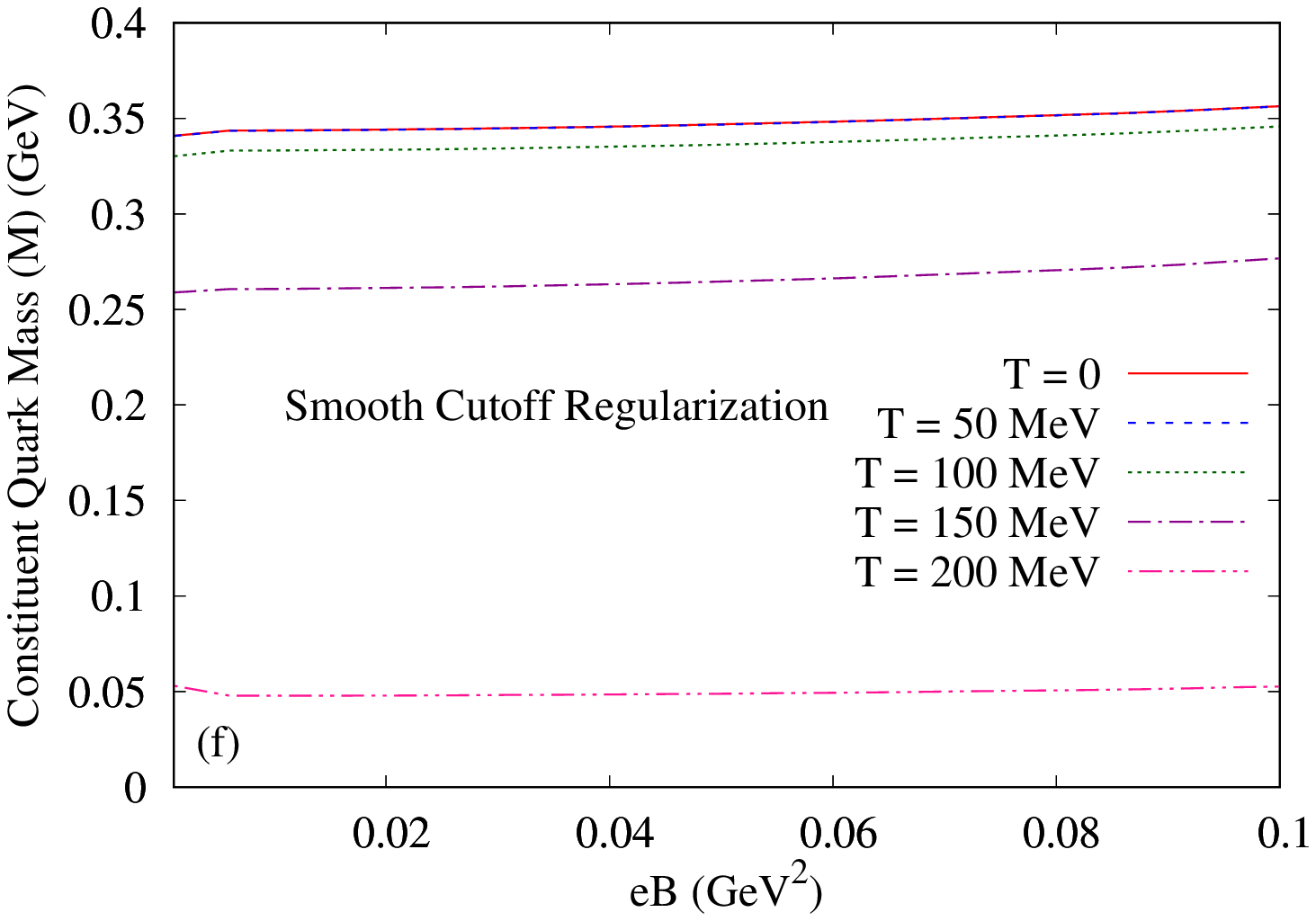}
	\end{center}
	\caption{The variation of (a) $\kappa_\text{u}$, (c) $\kappa_\text{d}$ and (e) $M$ as a function of temperature using the smooth cutoff regularization scheme. The variation of (b) $\kappa_\text{u}$, (d) $\kappa_\text{d}$ and (f) $M$ as a function of magnetic field using the smooth cutoff regularization scheme.}
	\label{fig.2}
\end{figure}

\section{NUMERICAL RESULTS \& DISCUSSIONS}\label{sec.results}
We begin this section by specifying the choice of parameters of the NJL model used in this work which are tabulated in Table~\ref{table1}.
The parameters are chosen so as to reproduce the phenomenological vacuum ($B=T=0$) values of quark condensate per 
flavor $\ensembleaverage{\psibar\psi}/N_f=-(243.5)^3$ MeV$^3$, pion decay constant $f_\pi=93$ MeV, pion mass $m_\pi=138$ MeV and the magnetic moment of the nucleons as
\begin{eqnarray}
\mu_\text{proton} \simeq 2.7928 ~\mu_\text{N} ~~~\text{and}~~~ \mu_\text{neutron}\simeq -1.9130 ~\mu_\text{N}.
\end{eqnarray}
The constituent mass $M$ and AMM of the quarks at $B=T=0$ comes out to be 
\begin{eqnarray}
M\simeq 343.8 ~\text{MeV} ~~~,~~~ \kappa_\tu\simeq 0.02399 ~\text{GeV}^{-1} ~~~\text{and}~~~ \kappa_\td\simeq 0.09595 ~\text{GeV}^{-1}.
\end{eqnarray}
Besides this, the chosen model parameters ensures that, at $B=T=0$, the relation $(F_2^\tu \sim F_2^\td)\simeq 0.05$ is satisfied which guarantees the isospin symmetry~\cite{Sadooghi,Bicudo:1998qb}.

We now show the numerical results for the AMM of the quarks ($\kappa_f$) and the constituent quark mass ($M$) which are obtained by solving 
the coupled gap Eqs.~\eqref{eq.ku}-\eqref{eq.gap.B} numerically. For all the numerical calculations, we have taken up to 1000 quark Landau levels. In Fig.~\ref{fig.1}, the variation of $\ku$, $\kd$ and $M$ as a function of temperature and external magnetic field are shown using the sharp cutoff regularization. We first notice from Figs.~\ref{fig.1}(a), (c) and (e), that both the AMM and constituent mass of the quarks are large in the chiral symmetry broken phase in the low temperature region. With the increase in temperature, they first remain almost unchanged up to a certain value of temperature and then fall rapidly around the pseudo-chiral phase transition temperature. Finally, at sufficiently high temperature region, the AMM of the quarks approach asymptotically to zero value whereas the constituent quark mass approaches the current quark mass value in the chiral symmetry restored phase. The temperature variation of the constituent quark mass in Fig.~\ref{fig.1}(e) is understandable physically from the fact that the mass gap $(M-m)$ is proportional to the chiral condensate $\Ensembleaverage{\psibar\psi}$, which is non-zero (zero) at the symmetry broken (restored) phase. Moreover, on comparing the finite magnetic field curves (blue, green and violet) with the zero magnetic field curve (red) in Fig.~\ref{fig.1}(e), we see that the external magnetic field strengthens the chiral condensate at all temperatures. This can also be noticed in Fig.~\ref{fig.1}(f) in which an overall increase of the constituent quark mass with the increase in magnetic field is seen.

On the other hand, the temperature dependence of the AMM of the quarks at zero magnetic field ( i.e. the red curves of Figs.~\ref{fig.1}(a), (c)) can be understood from Eqs.~\eqref{eq.F2.T} and \eqref{eq.kappa} where the AMM has a linear and dominant $M$ dependence apart from the complicated integral factor. Therefore, the temperature dependence of the AMM at zero magnetic field is almost similar to the temperature dependence of the constituent quark mass. Moreover, in the high temperature limit, the integral factor of Eq.~\eqref{eq.F2.T} approaches to zero and we get $\kappa_\tu \simeq \kappa_\td \simeq 0$. At non-zero external magnetic field, a linear $M$ dependence in the expression of AMM is not apparent from Eq.~\eqref{eq.If.1} unlike the zero magnetic field case and the RHS of Eq.~\eqref{eq.If.1} contains explicit AMM dependence as well as an implicit AMM dependence through the $M=M(\kappa)$. These complications forbid us to perform an analytical analysis of the finite magnetic field expressions. Nevertheless, numerically we see that, at sufficiently small value of external magnetic field, the complicated non-zero magnetic field expressions boil down to the exact zero-magnetic field results (compare the blue and red curves in Figs.~\ref{fig.1}(a), (c)); this is the consequence of the fact that at $B\simeq 0$, the Landau levels become infinitesimally close to each other and approach the continuum result of $B=0$.

Comparing Figs.~\ref{fig.1}(b) and (d), we see that AMM of the quarks are also slowly varying function of the external magnetic field alike the constituent quark mass depicted in Fig.~\ref{fig.1}(f). $\kappa_\tu$ decreases with the increase in external magnetic field whereas, $\kappa_\td$ shows an opposite trend. This is due to the opposite signs of the charges of up and down quarks producing an opposite response to the external magnetic field.

In Figs.~\ref{fig.1}(b), (d) and (f), we notice that both the AMM as well as the constituent quark mass suffer oscillations as the magnetic field changes. These oscillations appear to be an artifact of the use of sharp three-momentum regulator as commented in Refs.~\cite{Fukushima:2010fe,Miransky:2015ava}. We also notice that, the amplitudes of these oscillations are maximum around the pseudo-chiral phase transition region. Interestingly, it turns out that, when we use the smooth cutoff regularization scheme, these unphysical oscillations vanish as depicted in Fig.~\ref{fig.2}. Comparing Fig.~\ref{fig.1} and Fig.~\ref{fig.2}, we notice that, the oscillations appearing in Figs.~\ref{fig.1}(b), (d) and (f) get smeared out in Figs.~\ref{fig.2}(b), (d) and (f) keeping the overall qualitative and quantitative nature same. Moreover, the temperature dependence of AMM and constituent quark mass in Figs.~\ref{fig.1}(a), (c) and (e) suffers marginal change while switching to the smooth cutoff scheme as can be observed by comparing with Figs.~\ref{fig.2}(a), (c) and (e) respectively.


\section{SUMMARY \& CONCLUSION}\label{sec.summary}
In summary, using a gauged NJL model, we have evaluated the effective photon-quark-antiquark ($\gamma q \qbar$) vertex function in the mean field approximation. The lowest order diagram that contributes to the magnetic form factor and the AMM of the quarks is calculated at finite temperature in presence of arbitrary external magnetic field. The incorporation of finite temperature is done through the ITF of finite temperature field theory where the continuous energies of the loop particles are replaced with discrete Matsubara modes. The complete (including all the Landau levels) Schwinger propagator with non-zero AMM of the dressed quarks are considered while calculating the loop graphs. 
Using two different momentum cutoff (sharp and smooth) regularization scheme, we regularize the UV divergences arising from the vertex function and the parameters of our model are chosen to reproduce the well known phenomenological quantities at zero temperature and zero magnetic field, such as pion-decay constant ($f_\pi$), vacuum quark condensate, vacuum pion mass ($m_\pi$) as well as the magnetic moments of proton and neutron using CQM. Finally, the temperature as well as magnetic field dependence of the AMM and constituent quark mass are studied. 

Since the Schwinger propagator itself contains explicit AMM dependence, the magnetic form factor obtained from the vertex function in presence of magnetic field is also an explicit function of $\kappa$; along with, an implicit AMM dependence emerging from the constituent quark mass $M=M(\kappa)$. For this, the calculation of the AMM from the magnetic form factors requires solving a set of three \textit{coupled gap equations}.

We found that, the AMM as well as the constituent quark mass is large in the chiral symmetry broken phase in the low temperature region. Around the pseudo-chiral phase transition, $\kappa$ and $M$ suffer sudden decrease and at high temperature limit, both of them approach vanishingly small values at the symmetry restored phase. The value of $\kappa_\tu$ is seen to decrease slowly with the increase in magnetic field whereas an opposite trend is observed for $\kappa_\td$ due to the opposite sign of the charges of up and down quark. 
The oscillations seen in the magnetic field dependence of both the AMM and constituent mass of the quarks while using the sharp cutoff regularization scheme vanish when we use the smooth cut off regularization prescription.

\section*{Acknowledgments}
We acknowledge Dr. Arghya Mukherjee for useful discussions. SG is funded by the Department of Higher Education, Government of West Bengal. NC, SS and PR are funded by the Department of Atomic Energy (DAE), Government of India. 


\appendix

\section{CALCULATION OF MATRIX ELEMENTS} \label{app.mat.elem}
In this appendix, we will briefly sketch the derivation of the matrix elements $\expectationvalue{F}{\mcT \scrL_e(x)}{I}$ and 
$\expectationvalue{F}{\mcT \scrL_e(x) \scrL_G(y)}{I}$ leading to Eqs.~\eqref{eq.sfi.e} and \eqref{eq.sfi.eG}. 
The calculation of $\expectationvalue{F}{\mcT \scrL_e(x)}{I}$ is trivial since  
\begin{eqnarray}
\expectationvalue{F}{\mcT \scrL_e(x)}{I} = -|e|\Expectationvalue{ q(p;s,c,f)~ \qbar(p';s',c',f')}{ \mcT : \psibar(x)\hat{Q}\gamma^\mu \psi(x)\frac{}{} A_\mu(x):}
{\gamma(k;\lambda)} 
\end{eqnarray}
which on applying the Wick's theorem~\cite{Peskin:1995ev} becomes:
\begin{eqnarray}
\expectationvalue{F}{\mcT \scrL_e(x)}{I} &=& 
\wick[sep=3.2pt, offset=1.4em]{ -|e|\Expectationvalue{ \c1 q(p;s,c,f)~ \c2 \qbar(p';s',c',f')}{: \c1 \psibar(x)\hat{Q}\gamma^\mu \c2 \psi(x)\frac{}{} A_\mu(x):}{\gamma(k;\lambda)}
} \nn \\
&=& e^{ix\cdot(p+p'-k)}\ubar(p;s,c,f)  \FB{-|e|\hat{Q}\gamma^\mu \frac{}{}}_{c,c'}^{f,f'} v(p';s',c',f') \epsilon_\mu(k;\lambda).
\label{eq.sfi.e.1}
\end{eqnarray}

The evaluation of the quantity $\expectationvalue{F}{\mcT \scrL_e(x) \scrL_G(y)}{I}$ is bit involved. We have,
\begin{eqnarray}
\expectationvalue{F}{\mcT \scrL_e(x)\scrL_G(y)}{I} &=& -|e|G\Expectationvalue{ q(p;s,c,f)~ \qbar(p';s',c',f')}{ 
\mcT : \psibar(x)\hat{Q}\gamma^\mu \psi(x)\frac{}{} A_\mu(x)::\psibar(y)\psi(y) \psibar(y)\psi(y):}{\gamma(k;\lambda)} \nn \\
 && \hspace{-2cm}  + |e|G\Expectationvalue{ q(p;s,c,f)~ \qbar(p';s',c',f')}{ 
	\mcT : \psibar(x)\hat{Q}\gamma^\mu \psi(x)\frac{}{} A_\mu(x)::\psibar(y)\gamma^5\tau^i\psi(y) \psibar(y)\gamma^5\tau^i\psi(y):}{\gamma(k;\lambda)}.
\end{eqnarray}
Applying Wick's theorem, we obtain
\begin{eqnarray}
\expectationvalue{F}{\mcT \scrL_e(x)\scrL_G(y)}{I} = 
\wick[sep=3.2pt, offset=1.4em]{-|e|G \Big\langle \c1 q(p;s,c,f)~ \c2 \qbar(p';s',c',f')\Big| 
	 : \c3 \psibar(x)\hat{Q}\gamma^\mu \c4 \psi(x)\frac{}{} \c5 A_\mu(x) \c4 \psibar(y) \c2 \psi(y) \c1 \psibar(y) \c3 \psi(y): \Big| \c5 \gamma (k;\lambda) \Big\rangle
 } \nn \\
\wick[sep=3.2pt, offset=1.4em]{-|e|G \Big\langle \c1 q(p;s,c,f)~ \c2 \qbar(p';s',c',f')\Big| 
	: \c3 \psibar(x)\hat{Q}\gamma^\mu \c4 \psi(x)\frac{}{} \c5 A_\mu(x) \c1 \psibar(y) \c3 \psi(y) \c4 \psibar(y) \c2 \psi(y): \Big| \c5 \gamma (k;\lambda) \Big\rangle
} \nn \\
\wick[sep=3.2pt, offset=1.4em]{+|e|G \Big\langle \c1 q(p;s,c,f)~ \c2 \qbar(p';s',c',f')\Big| 
	: \c3 \psibar(x)\hat{Q}\gamma^\mu \c4 \psi(x)\frac{}{} \c5 A_\mu(x) \c4 \psibar(y)\gamma^5\tau^i \c2 \psi(y) \c1 \psibar(y)\gamma^5\tau^i \c3 \psi(y): \Big| \c5 \gamma (k;\lambda) \Big\rangle
} \nn \\
\wick[sep=3.2pt, offset=1.4em]{+|e|G \Big\langle \c1 q(p;s,c,f)~ \c2 \qbar(p';s',c',f')\Big| 
	: \c3 \psibar(x)\hat{Q}\gamma^\mu \c4 \psi(x)\frac{}{} \c5 A_\mu(x) \c1 \psibar(y)\gamma^5\tau^i \c3 \psi(y) \c4 \psibar(y)\gamma^5\tau^i \c2 \psi(y): \Big| \c5 \gamma (k;\lambda) \Big\rangle
} + \cdots 
\end{eqnarray}
where we have omitted few other possible contractions as they do not contribute to the electromagnetic form factors of the quarks.  
Simplification of the above expression yields
\begin{eqnarray}
\expectationvalue{F}{\mcT \scrL_e(x) \scrL_G(y)}{I} &=& e^{iy\cdot(p+p')} e^{-ix\cdot k} \ubar(p;s,c,f)  
\FB{-2GS(y,x)|e|\hat{Q}\gamma^\mu S(x,y) \right. \nn \\
	&& \left. + 2G\gamma^5\tau^i S(y,x)|e|\hat{Q}\gamma^\mu S(x,y) \gamma^5\tau^i \frac{}{}}_{c,c'}^{f,f'} v(p';s',c',f') \epsilon_\mu(k;\lambda)
+ \cdots
\label{eq.sfi.eG.1}
\end{eqnarray}

\bibliographystyle{apsrev4-1}
\bibliography{snigdha}

\end{document}